\documentclass[twocolumn,showpacs,preprintnumbers,amsmath,amssymb,prl]{revtex4}

\usepackage{amsmath,amssymb}
\usepackage{graphicx}

\usepackage[dvips,colorlinks=true,bookmarks=false,citecolor=blue,urlcolor=blue]{hyperref} %latex w/dvips

\begin{document}

\title{Engineered Optical Nonlocality in Nanostructured Metamaterials}

\author{Alexey A. Orlov$^{1}$, Pavel M. Voroshilov$^{1}$, Pavel A. Belov$^{1,2}$, Yuri S. Kivshar$^{1,3}$}
\affiliation{$^{1}$St. Petersburg University of Information Technologies, Mechanics and Optics (ITMO), St. Petersburg 197101, Russia \\
$^{2}$Queen Mary University of London, Mile End Road, London E1 4NS, UK \\
$^{3}$Nonlinear Physics Centre, Research School of Physics and Engineering, Australian National University, Canberra ACT 0200, Australia}
%\email{alexey.orlov@phoi.ifmo.ru}

\begin{abstract}
We analyze dispersion properties of metal-dielectric nanostructured metamaterials. We demonstrate that, in a sharp contrast to
the results for the corresponding effective medium, the structure demonstrates strong {\em optical nonlocality} due to excitation of
surface plasmon polaritons that can be engineered by changing a ratio between the thicknesses of metal and dielectric layers.
In particular, this nonlocality allows the existence of an additional extraordinary wave that manifests itself in the splitting of the
TM-polarized beam scattered at an air-metamaterial interface.
\end{abstract}


\maketitle

It is well accepted that the properties of optical composites and nanostructured metamaterials can be described by effective parameters
derived in the limit when structural elements are much smaller than the wavelength~\cite{smith_josab}.  The effective medium is an important
concept of the homogenization theory based on field averaging, and it provides a physical insight into the optical response of complex micro- and nanostructured media being also useful for different types of waves~\cite{book}. However, it was already established that the effective medium models do not provide a complete information, and they should be corrected in some cases, e.g. in the recently analyzed case of plasmonic nanorod metamaterials in the epsilon-near-zero regime, where the performance of such structures is affected by nonlocal response~\cite{pollard_prl}.

 Here we demonstrate that the dispersion properties of metal-dielectric periodic nanostructured metamaterials are dramatically affected by a nonlocal response due to excitation and coupling of surface plasmon polaritons at the metal-dielectric interfaces, so in many cases the effective medium approach fails to describe correctly the optical properties of such structures. The difference is dramatic, and it can't be taken into account by small corrections~\cite{pollard_prl,elser_apl}. Nevertheless, we reveal that the strength of optical nonlocality can be engineered in a rather simple way,
just changing a ratio between the thicknesses of metal and dielectric layers, so that the medium with equal layers demonstrates weak nonlocality.

The electromagnetic response of periodic layered metal-dielectric nanostructures has been a subject of many theoretical and experimental studies. Such structures represent the simplest nanostructured metamaterials, they were suggested for a number of applications including superlenses with subwavelength resolution~\cite{Pendry-Ramakrishna,Belov,xuanli}, as a simple realization of the so-called hyperlens~\cite{HyperScience}, as well as for nanolithography~\cite{xiong}, optical nanocircuits~\cite{metactronics}, invisibility cloaks~\cite{cloaking}, and even three-dimensional negative refraction~\cite{verhagen_prl}. In many cases, the effective medium is a conventional approach for describing optical properties of such structures, and it allows to introduce the effective permittivities and show that such structures behave as uniaxial metamaterials with the optical axis parallel to the layers.

In this Letter we reveal that, depending on the thicknesses of layers, the metal-dielectric structures can demonstrate strong
optical nonlocality due to the excitation of surface plasmon polaritons. In particular, we predict the existence of an additional extraordinary
wave which affect dramatically the scattering of the TM-polarized wave at an air-metamaterial interface generating two waves with negative and positive refraction, respectively.

\begin{figure}
\includegraphics[width=0.5\textwidth]{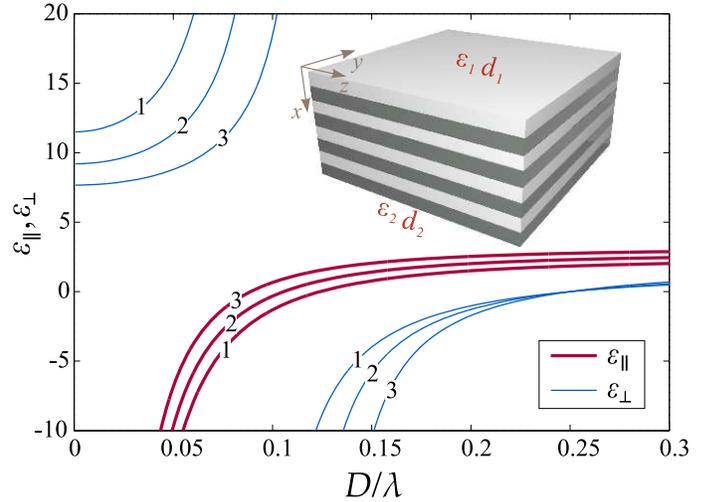}
\caption{(Color online) Effective medium model. Permittivity tensor components as functions of normalized frequency for three types of periodic
layered metal-dielectric nanostructures with the fixed lattice spacing $D=d_1+d_2$, but formed by pairs of metal and dielectric layers
with the thicknesses (1) $d_2=1.5d_1$, (2)~$d_1=d_2$, and (3) $d_1=1.5d_2$, respectively.}
\label{fig1}
\end{figure}

\begin{figure*}
  \includegraphics[width=\textwidth]{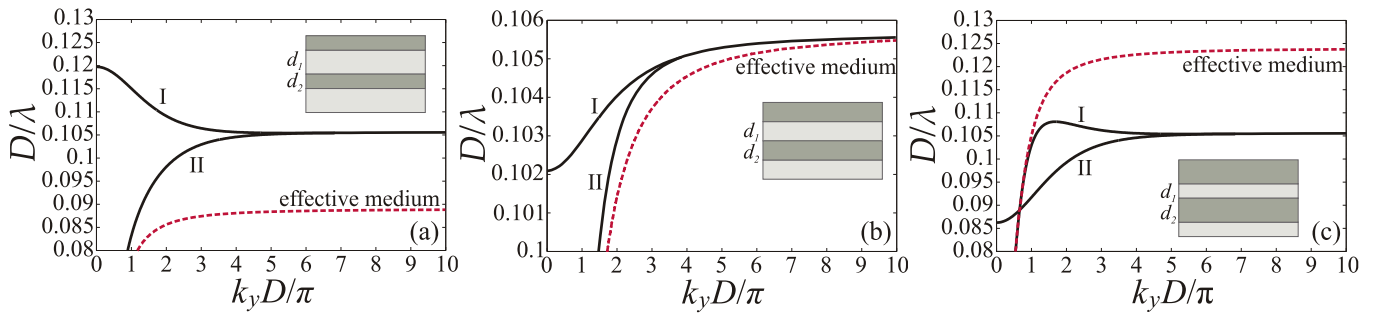}
\caption{(Color online) Dispersion diagrams and the geometries of the periodic nanostructures composed of alternating metal and dielectric layers.
Thicknesses of the layers are: (a) $d_2=1.5d_1$, (b)~$d_1=d_2$, and (c) $d_1=1.5d_2$. Two dispersion branches corresponding to the actual structures are numbered by Roman numerals. }
\label{fig:dispdiags}
\end{figure*}

We consider a nanostructured metamaterial formed by a periodic layered structure (see the insert in Fig.~\ref{fig1}), where metal and dielectric layers have different thicknesses. When the wavelength of radiation is much larger than the thickness of any layer,  it is usually assumed that the effective medium approach is valid and the permittivity tensor for an uniaxial anisotropic medium has the following form:
\begin{equation}
\varepsilon_\mathrm{eff} = \left( \begin{matrix}
\varepsilon_\bot & 0 & 0\\0 & \varepsilon_\| & 0\\0 & 0 & \varepsilon_\|
\end{matrix}\right), \
\begin{array}{lcl}
\varepsilon_\| = \dfrac{\varepsilon_1 d_1 + \varepsilon_2 d_2}{d_1 + d_2},
\\
\varepsilon_\bot = \dfrac{\varepsilon_1 \varepsilon_2 (d_1 + d_2)}{\varepsilon_2 d_1 + \varepsilon_1 d_2},
\end{array}
 \label{eq:diel-tens}
\end{equation}
where $\varepsilon_1, \varepsilon_2$ and $d_1, d_2$ are the dielectric permittivities and thicknesses of the layers, respectively.

A dispersion equation for extraordinary waves supported by the such an effective uniaxial medium and propagating in the $xy$-plane has the form
\begin{equation}
\frac{k_x^2}{\varepsilon_\|} +  \frac{k_y^2}{\varepsilon_\bot} = \left( \frac{\omega}{c} \right)^2,
\label{eq:disp-cryst}
\end{equation}
and it establishes a relation between the wavevector components $k_x$, $k_y$ and the frequency $\omega$. Below the plasma frequency, the permittivities of metal and dielectric have different signs, and this ensures a key feature of metal-dielectric nanostructures (MDNs):
 the principal elements of their permittivity tensor can have nearly arbitrary values.
For example, if $\varepsilon_1=-\varepsilon_2 d_1/d_2$ then the structure has very large permittivity ($\varepsilon_\perp \rightarrow \infty$),
 and it represents realization of an ``epsilon-very-large'' (EVL) material \cite{HIR}. If $\varepsilon_1=-\varepsilon_2 d_2/d_1$ then an MDN has near zero permittivity ($\varepsilon_\|=0$) and is a good candidate for realization of the so-called ``epsilon-near-zero'' (ENZ) material~\cite{squeezing}.
Basically, the EVL and ENZ materials are optical conductors and insulators which form a basis of metactronics \cite{metactronics}.
If $\varepsilon_\|$ and $\varepsilon_\bot$ have different signs then the MDN is a typical realization of the so-called indefinite medium \cite{smith-schurig}.

In this work we considered three MDNs formed by layers of metal and dielectric with various thickness ratios, but fixed total period. Configurations of the structures are illustrated schematically in the insets of Figs.~\ref{fig:dispdiags}(a-c). The dispersion diagrams $D/\lambda(k_y)$ for the three MDNs under consideration shown in Figs.~\ref{fig:dispdiags}(a-c) were computed using two approaches: the effective medium model (approximate approach) and the well-known classical dispersion relation for one-dimensional photonic crystals (exact description).

The properties of such MDNs are investigated at wavelengths which are approximately 10 times larger than the structure period. The permittivity of dielectric is assumed to be constant and equal to $\varepsilon_1 = 4.6$. Permittivity of metal is given by the Drude model: $\varepsilon_2 = 1 - \lambda^2/\lambda_p^2$, where $\lambda_p = 4 D$. The losses in metal are neglected in order to simplify the consideration. The frequency dependencies of $\varepsilon_\|$ and $\varepsilon_\perp$ for three types of MDNs with different ratios of layer thicknesses are shown in Fig.~\ref{fig1}. The singularities of $\varepsilon_\perp$  and nulls of $\varepsilon_\|$ corresponding to EVL and ENZ behavior, respectively, are clearly visible.

The information about dispersion properties of MDN provided by effective medium model is approximate, and it does not take into account actual periodicity of the structure. The general expressions for  $\varepsilon_\|$ and $\varepsilon_\perp$ in Eq.(\ref{eq:diel-tens}) are deduced under quasi-static approximation by using an assumption that the electric field does not vary inside of the layers. If permittivities of the constituent layers are positive then the requirement that the period of the structure (and thus the thicknesses of the constituent layers) is much smaller than the wavelength of operation
ensures negligibly small variation of fields since spatial harmonics propagating inside of the layers have small wave numbers.
However, such a reasoning is not valid if materials with negative permittivity (e.g. metals) are involved because the interfaces between materials with positive and negative permittivity support surface plasmon polaritons (SPPs) which form an alternative channel for wave propagation and may cause significant field variation in the layers. The SPPs at the individual interfaces of an MDN exist at the frequencies below the SPP resonance which corresponds to condition $\varepsilon_1=-\varepsilon_2$. It is logical to expect violation of classical homogenization requirement at these frequencies, but only for oblique propagation of waves through the MDN since the waves propagating across the layers cannot couple to SPPs. Note, that the SPP resonance condition, which corresponds to the strongest confinement of SPPs and thus leads to strongest variation of fields across the layers, connects only permittivities of the materials which form the interfaces where SPPs propagate, and it does not depend on the thicknesses of the layers.

The dispersion characteristics of actual periodic MDN can be evaluated from Maxwell's equations with imposed periodic boundary conditions.
A consideration of the MDN as one-dimensional photonic crystal leads to the following exact dispersion relation for eigenmodes with transverse magnetic (TM) polarization:
\begin{widetext}
\begin{align}
\cos(k_x D) = \cos(k_x^{(1)} d_1)\cos(k_x^{(2)} d_2) -
\frac{1}{2} \left( \frac{\varepsilon_2 k_x^{(1)}}{\varepsilon_1 k_x^{(2)}} + \frac{\varepsilon_1 k_x^{(2)}}
{\varepsilon_2 k_x^{(1)}} \right) \sin(k_x^{(1)} d_1)\sin(k_x^{(2)} d_2)
\label{eq:disp-matr} ,
\end{align}
where $k_x^{(1,2)}=[\varepsilon_{1,2}(\omega/c)^2-k_y^2]^{1/2}$ are $x$-components of wavevectors in the corresponding layers.
\end{widetext}

The MDNs with 3 different ratios of layers thicknesses (3/2, 1 and 2/3, respectively) were chosen in order to demonstrate different behaviors of dispersion curves.
In the first case, the SPP resonance (in our case corresponding to $D/\lambda=0.1055$) lays over the frequency where EVL behavior is expected ($D/\lambda=0.089$).
In the second case, the frequencies are chosen to be equal. In the third case, the SPP resonance appears well below EVL frequency ($D/\lambda=0.124$).
In all cases the dispersion curves consists of two branches and have SPP resonance as asymptote if $k_y \to \infty$.
However, in the 1st case the upper branch corresponds to a backward wave while the second branch doesn't. In the 2nd case both branches correspond to forward waves
and co-exist at the same frequency range in contrary to the 1st case.
In the 3rd case the branches cross each other right at the frequency corresponding to ENZ behavior of MDN ($D/\lambda=0.089$) and one of the branches has a maximum leading to existence of backward and forward waves simultaneously at the same range of frequencies.
The effective medium model in all cases predicts only one propagating wave at all frequencies.
The presence of two propagating waves is a consequence of nonlocality and strong spatial dispersion which are caused by SPPs at the interfaces of the layers.

We have verified that the behavior observed for MDNs with the period 10 times smaller than wavelength remains qualitatively the same for MDNs with 2, 4 and 8 times smaller periods (i.e. 20, 40 and 80 times smaller than wavelength). The convergence to an asymptotic line corresponding to SPP resonance becomes weaker (the branches reaches this line at larger $k_y$), but the topology of the branches remains the same. On the other hand, the decrease of period pushes the dispersion curves closer to those predicted by the effective-medium approximations for small $k_y$. It is interesting that this convergence also depends on the ratio of the layers thicknesses. In the case of equal layers the convergence is faster than for different layers. However, even for very small period the correspondence between the results of the effective model and actual dispersion properties of MDNs is not satisfactory, especially for the waves with relatively large $k_y$ which are mainly used in the problem of subwavelength imaging.

\begin{figure}
  \includegraphics[width=0.5\textwidth]{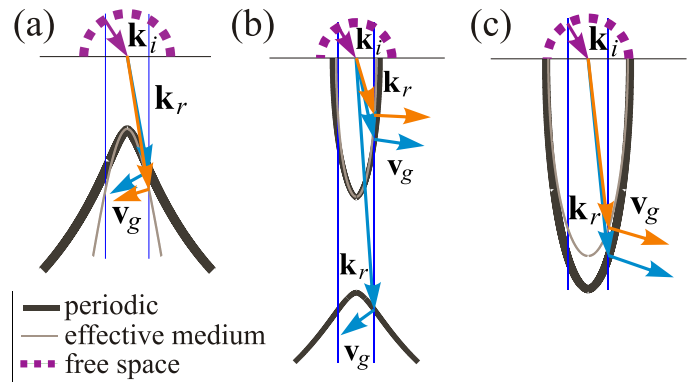}
\caption{\label{fig:isorefr}
(Color online) Diagrams of beam refraction at the air-MDN interface plotted both for local effective medium model and actual MDN with help of isofrequency contours at three different frequencies: a) $D/\lambda = 0.085$, (b) $D/\lambda = 0.094$, and (c)~$D/\lambda = 0.106$.}
\end{figure}

\begin{figure*}
  \includegraphics[width=\textwidth]{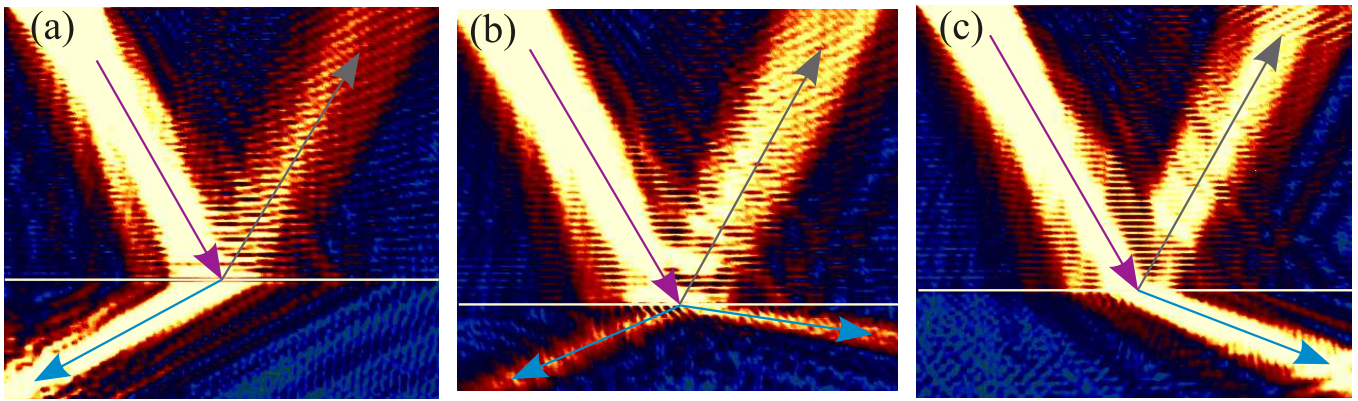}
\caption{\label{fig:refr}
(Color online) Numerical studies of the beam refraction in the three cases corresponding to Figs.~\ref{fig:isorefr}(a-c), respectively.}
\end{figure*}

The isofrequency contours can be used for describing the beam refraction at the interfaces separating free space and the layered structure. They allow to understand the character of refracted waves and find directions of their phase and group velocities.
For the three typical cases of isofrequency contours, the refraction diagrams are plotted in Fig.~\ref{fig:isorefr}. In the first case [see Fig.~\ref{fig:isorefr}(a)], there exists only one negatively refracted beam, both for the effective medium model and transfer-matrix method. Here we have a situation where the effective medium model can be applicable. The same situation is demonstrated in Fig.~\ref{fig:isorefr}(c), but one positively refracted beam is observed. The simultaneous appearance of ellipse and hyperbolic-like contour in Fig.~\ref{fig:isorefr}(b) leads to birefringence phenomena in the metal-dielectric structure. In this case both negatively and positively refracted TM-polarized beams should appear simultaneously.

In order to verify this prediction we have performed numerical simulations with a commercial full-wave electromagnetic solvers package~\cite{cst} for the wave refraction at an interface at different frequencies [see Figs.~\ref{fig:refr}(a-c)]. It should be noted that such simulations are quite hard to accomplish since the beam width is equal to about 10 wavelengths of radiation $\lambda$, with the smallest mesh element reaching $\lambda/500$.  At $D/\lambda = 0.085$ only one negatively refracted wave is observed [Fig.~\ref{fig:refr}(a)] and the effective medium model can be applicable. Appearance of an ellipse in the isofrequency contours at $D/\lambda = 0.094$ shown in Fig.~\ref{fig:isorefr}(b) leads to birefringence [Fig.~\ref{fig:refr}(b)]. Negatively refracted beam still appears but in addition to it a positively refracted beam is observed. In this case, the effective medium model is not applicable, since it does not predict the presence of the two waves describing only one of them. At the higher frequency ($D/\lambda = 0.106$) only one positively refracted ray is observed, and this fact is well described by the effective medium model.

Our results call for deeper studies of the applicability limits of the effective medium models, and they may be useful to explain experimental results in more complex settings. In particular,  a few years ago Kozyrev et al.~\cite{ilya} observed experimentally the waveform splitting into two or multiple beams after the propagation of electromagnetic waves through a slab of a magnetic metamaterial. Such effects can not be described by an effective-medium model being characterized by some effective permittivity and permeability. Instead,  the observed transmission properties of metamaterials are affected significantly by the internal structure, such that the multiple beam formation observed in the experiments can be attributed to the excitation of magnetoinductive waves.

In conclusion,  we have studied the dispersion properties of periodic metal-dielectric structures by employing the exact transfer-matrix approach and an effective medium model.  We have revealed a substantial difference between the results provided by these two approaches, even in the limit when the structure spacing is much smaller than the radiation wavelength. These result question the applicability of the averaged parameters and effective media in the cases when the structure support surface waves. In particular, we have found two dispersion branches of extraordinary waves instead of one branch predicted by the effective medium model. Importantly, the effective optical nonlocality has been shown to depend dramatically on a ratio between the thicknesses of metal and dielectric layers, and it can be engineered by changing this ratio, with the weakest nonlocality in the case of equal thicknesses. Strong nonlocal effects predicted here can be observed as the splitting of the TM-polarized wave at an interface between air and metal-dielectric nanostructured metamaterial into two waves with positive and negative refraction. Our results suggest the presence of strong spatial dispersion in many types of nanostructured materials and, in particular, they provide a proof that metal-dielectric nanostructures are inherently nonlocal metamaterials.

The authors acknowledge a support from the Ministry of Education and Science of Russian Federation (Russia), EPSRC (UK), and Australian Research Council (Australia).

\end{document}